\documentclass{article}

\begin{document}

\newtheorem{defn}{Definition}
\def\thedefn{\thesection.\arabic{defn}}
             
\newtheorem{teor}[defn]{Theorem}
\newtheorem{ejem}[defn]{Example}
\newtheorem{lema}[defn]{Lemma}
\newtheorem{rema}[defn]{Remark}
\newtheorem{coro}[defn]{Corollary}
\newtheorem{prop}[defn]{Proposition}

\makeatother
\font\ddpp=msbm10  at 11 truept 
\def\R{\hbox{\ddpp R}}     
\def\C{\hbox{\ddpp C}}     
\def\L{\hbox{\ddpp L}}    
\def\S{\hbox{\ddpp S}}
\def\Z{\hbox{\ddpp Z}}
\def\Q{\hbox{\ddpp Q}}     
\def\N{\hbox{\ddpp N}}

\newcommand{\dps}{d_{\psi}}
\newcommand{\df}{d_{\phi}}

\newcommand{\M}{{\cal M}}
\newcommand{\Mo}{{\cal M}_0}
\newcommand{\be}{\begin{equation}}
\newcommand{\ee}{\end{equation}}
\newcommand{\la}{\Lambda}
\newcommand{\dem}{\noindent {\rm Proof: }}
\newcommand{\inte}{\int_{0}^{1}}
\newcommand{\gam}{\gamma}
\newcommand{\eps}{\epsilon}
\newcommand{\<}{\langle}
\renewcommand{\>}{\rangle}
\newcommand{\Om}{\Omega^1}
\renewcommand{\(}{\left(}
\renewcommand{\)}{\right)}
\renewcommand{\[}{\left[}
\renewcommand{\]}{\right]}
\newcommand{\om}{\omega}
\newcommand{\me}{\frac{1}{2}}
\newcommand{\Mt}{\widetilde{\M}}
\newcommand{\cat}{{\mathop{\rm cat}\nolimits}}

\newcommand{\cvd}{{\rule{0.5em}{0.5em}}\smallskip}

\hyphenation{Lo-rent-zian}

\title{{\bf On smooth Cauchy hypersurfaces and  Geroch's splitting theorem}}
\author{Antonio N. Bernal,  and Miguel S\'anchez\thanks{The authors acknowledge Prof. P. Ehrlich's clarifying comments  on Geroch's theorem and subsequent references.
The second-named author has been partially supported by a  MCyT-FEDER Grant BFM2001-2871-C04-01.
}}
\maketitle

\begin{center}
{\small Dpto. de Geometr\'{\i}a y Topolog\'{\i}a, Facultad de Ciencias, Fuentenueva s/n, E--18071 Granada, Spain}
\end{center}

\vspace*{.4cm}

\begin{flushright} 
{\em To Professor J.K. Beem, \\
in the year of his retirement}
\end{flushright} 

\vspace*{.2cm}

\begin{center}
Abstract
\end{center}

\noindent Given a globally hyperbolic spacetime $M$, we show the existence of a {\em smooth} spacelike Cauchy hypersurface $S$ and, thus, a global diffeomorphism between $M$ and $\R \times S$. 

\smallskip

\section{Introduction}

In a classical article published in 1970, Geroch \cite{G} proved the equivalence, for a spacetime, between global hyperbolicity and the existence of a Cauchy hypersurface $S$. As he stated clearly, the results were obtained at a topological level. In fact, he proved the existence of a {\em continuous} time function $t: M \rightarrow \R$ such that each level $t=$constant is a ({\em topological}) Cauchy hypersurface and, then, $M$ is {\em homeomorphic} to $\R \times S$. The improvement of these topological results in smooth differentiable ones is important not only as a typical mathematical challenge but also from the theoretical viewpoint: Cauchy hypersurfaces are the natural subsets where initial conditions to differential equations (as Einstein's equation) are posed. Moreover, any achronal hypersurface $N$ can be seen as a graph on a Cauchy hypersurface; this allows to study several properties which involve the differentiable structure  of $N$ (as mean curvature),  {\em provided that the splitting is smooth}; see for example, \cite{EH}.

In general, continuous maps between smooth manifolds can be approximated by smooth maps. Thus, it is reasonable to expect that, with some effort, one could fill the details to strengthen the topological results. Some authors have claimed that function $t$ (and, thus, the Cauchy hypersurface) can be smoothed and, therefore, $M$ is diffeomorphic to $\R \times S$; for example, see the end of the proof of Proposition 6.6.8 in the very influential book by Hawking and Ellis \cite{HE} (or the more recent book \cite[p. 209]{Wa}). In fact, this has been assumed in very different contexts where smoothness seems unavoidable 
(see, for example, \cite{F}, \cite[Chapter 8]{Ma} or \cite{Uh}). 

Nevertheless, as far as we know, general smoothing procedures have resisted the 
attempts of formalisation. Budic and Sachs \cite{BS} proved $C^1$-smoothing of time functions for {\em deterministic} globally hyperbolic spacetimes. Then, Seifert  
 \cite{Se} claimed the existence of a general procedure for smoothing time functions. But his proof is complicated and seems unclear; no cleaner version of this result has been published since then. 
 Later on, Dieckmann 
 \cite{D2, D} 
clarifies some points in Geroch's proof, but cites Seifert at the crucial point for smoothing (see \cite[proof of Theorem]{D2}). In general, in spite of these two references, most specialists in pure Lorentzian Geometry {\em do  not} affirm that a smooth Cauchy hypersurface must exist, even when the context could suggest it (say, expressions such as ``consider a globally hyperbolic spacetime with a smooth Cauchy hypersurface $S$'' are commonly used, when necessary). For example, among our  references clearly posterior to Seifert's article, see \cite{BE, BEE,  EH, Ga,  Ne, O}. 

Summing up,  Sachs and Wu \cite[p. 1155]{SW} claimed in 1977: 
\begin{quote}
...This is one of the folk theorems of the subject. It is not difficult to prove that every Cauchy surface is in fact a Lipschitzian hypersurface in $M$ \cite{Pe}. However, to our knowledge, an elegant proof that his Lipschitzian submanifold can be smoothed out to such an $N$ above is still missing.
\end{quote}
This ``folk question'' is regarded as open in the first edition of the classical book by Beem and Ehrlich \cite[p. 31]{BE}, and remains open in the second edition of 1996,
with Easley \cite[p. 65]{BEE}. As far as we know, no formalisation of the quoted result has been obtained since then.
On the other hand, it is worth mentioning that some properties of volume functions (which appear in Geroch's proof) have been studied in  \cite{D} (see also \cite[pp. 65--69]{BEE}), and some smooth splitting results in different contexts have been obtained, see for example, \cite{Ne}, \cite[Chapter 14]{BEE}.  

The aim of the present article is to give a simple, self-contained and detailed proof of the following result:

\begin{teor} \label{tsuave}
Any globally hyperbolic spacetime admits a smooth spacelike Cauchy hypersurface $S_0$ and, then, it is diffeomorphic to $\R \times S_0$.
\end{teor}
This paper is organised as follows. Among the preliminaries in Section \ref{s2},  we state what can be asserted from Geroch's splitting theorem, Lemma  \ref{ro}, Proposition \ref{p0}. In Section \ref{s3} some technical properties of Cauchy hypersurfaces are proven. We do not try to give general properties here; plainly,  we give direct proofs to  the results needed later, for the sake of clarity and completeness. We prove, essentially (compare with \cite{BILY}, \cite{Ga}):  any closed spacelike hypersurface $N$ which lies in one side of a Cauchy hypersurface is  achronal (Proposition \ref{l1}) and, thus, if it lies between two Cauchy hypersurfaces, then it is a Cauchy hypersurface too (Corollary \ref{p1}). The necessity of the hypotheses is discussed in Remarks \ref{re1},
\ref{re2}. 
In Section \ref{s4} we prove Theorem \ref{tsuave}, according to the following two steps. Fix two Cauchy hypersurfaces, one of them, $S_{1}$, in the past of the other one, $S_{2}$. Roughly: 

(1) For each point $p\in S_2$ a smooth function $h_p$ with {\em compact support} can be constructed such that $\nabla h_p$ is timelike (or 0) in the past of $S_2$ (Lemma \ref{local}). This function is constructed from the square of time-separation, which is either null or a quadratic polinomial in normal coordinates and, thus, essentially smooth (see formula (\ref{hp})). 

(2) By using technical properties related to paracompactness (Lemma \ref{topol}), a locally finite set of these functions can be summed in such a way that the sum, $h$, restricted to 
$J^+(S_{1}) \cap J^-(S_{2})$, admits regular level hypersurfaces  which are spacelike and 
do not touch each
$S_{i}$ (Proposition \ref{global}). 

\section{Preliminaries. Geroch's result} \label{s2}

We will follow usual conventions in Lorentzian Geometry as in \cite{BEE}, \cite{G}, \cite{O},  \cite{Pe}, \cite{SW}. 
In particular, a {\em spacetime} $M$ will be a connected $n-$manifold $n\geq 2$ 
endowed with a time-oriented Lorentzian metric $g$. Differentiability $C^k$, $k \in 
\N \cup \{ \infty\}$, ($\N =\{1, 2, \dots \}$), will be assumed for both the manifold and the metric,  and the term ``smooth'' will mean ``$C^k$-differentiable''. Sometimes an auxiliary  complete Riemannian metric $g_R$ (with associated norm $\parallel \cdot \parallel_R$) will be needed. Recall that  the existence of a Riemannian metric on any paracompact manifold (as those admitting a Lorentzian metric, see the discussion above Lemma \ref{topol})
is well-known; it can be chosen complete by using Whitney's embedding theorem (moreover, any Riemannian metric admits a conformal metric which is complete
\cite{NO}). 

$M$ is {\em globally hyperbolic} if it is strongly causal and $J^+(p)\cap J^-(q)$ is compact for any $p, q \in M$. In this case, it is not hard to prove that, for any two compact subsets $K_1, K_2$ the set
\be \label{jk}
K= J^+(K_1) \cap J^-(K_2)
\ee
is compact too.
A {\em hypersurface} $H$ in $M$ is a embedded topological $(n-1)$-submanifold without boundary. $H$ can be regarded as a subset of $M$ and, then, $H$ will be {\em closed} if it is a closed subset of $M$. A {\em spacelike} hypersurface is a embedded $C^r$-hypersurface ($r\in \{1,\dots k\}$) such that its tangent space at each point is spacelike.
A {\em Cauchy hypersurface} in $M$ is a subset $S$ that is met exactly once by every inextendible timelike curve in $M$. Then, $S$ will be a closed achronal  
connected (topological) hypersurface and it is intersected by any inextendible causal curve \cite[Lemma 14.29]{O} (the intersection may be a closed geodesic segment instead a single point, if the curve is lightlike there).

The following result should be well-known (see for example, \cite[p. 444, Property 7]{G}, \cite[Proposition 14.31]{O}), even though we include its proof for the sake of completeness and further referencing.

\begin{lema} \label{ro}
Let $M$ be a ($C^k$-)spacetime which admits a $C^r$-Cauchy hypersurface $S$, $r\in \{0, 1, \dots k\}$. Then  $M$ is $C^r$-diffeomorphic to $\R \times S$ and all the $C^r$-Cauchy hypersurfaces are $C^r$ diffeomorphic.
\end{lema}
\dem It is well-known that any spacetime admits a smooth timelike vector field $T$. Moreover, $T$ can be assumed to be complete (otherwise, choose any auxiliary complete Riemannian metric 
$g_R$ and take the complete vector field $T/\parallel T \parallel _R$). Let $\phi$ be the flow of $T$ and consider the map:
$$ \Phi: \R \times S \rightarrow M \quad \quad (s,x) \rightarrow \phi_s(x). $$
As $S$ is a Cauchy hypersurface, then $\Phi$ is bijective and, by construction, $\Phi$ is a $C^r$ diffeomorphism (in the case $r=0$, $\Phi$ is a homeomorphism -use the classical Brouwer theorem on the invariance of the domain). 

Let $S'$ be any other $C^r$-Cauchy hypersurface. Putting 
\be \label{defro}
\Phi^{-1}(z)=(s(z), \rho(z)),
\ee
 it is clear that the map $ S' \rightarrow S , \; z \rightarrow \rho(z) $
is a $C^r$-diffeomorphism.
\cvd

\begin{rema}
{\rm We emphasize that, in this result, each hypersurface at constant $s$ is not necessarily a Cauchy hypersurface.}
\end{rema}
\noindent Geroch proved in \cite{G} (see Section 5, Theorem 11, plus footnote 26):
\begin{prop}\label{l0} Assume that the spacetime $M$ is globally hyperbolic. Then there exists a continuous and onto map $t: M \rightarrow \R $ satisfying:

    (1) $S_a := t^{-1}(a)$ is a Cauchy hypersurface, for all $a\in \R$.

    (2) $t$ is strictly increasing on any causal curve.
\end{prop}
\noindent In the proof of this result,  $t$ is obtained by a  famous argument: essentially,  choose a measure on $M$ with finite total volume and put 
\be \label{tiempo}
t(z)= {\rm ln} \left( {\rm vol}(J^-(z))/{\rm vol}(J^+(z))\right) 
\ee
(see also Theorem 3.26 in \cite{BEE} as well as pp. 65--72, for a discussion about the admissible measures {\em \`a la Dieckmann} \cite{D}). Then, as a consequence one has: 

\begin{prop} \label{p0}
Let $M$ be a globally hyperbolic spacetime and $S$ one of its Cauchy hypersurfaces.
Then there exist a homeomorphism
\be \label{ed1}
\Psi: M\rightarrow \R \times S , \quad z \rightarrow (t(z), \rho(z)),
\ee
which satisfies:

(a) Each level hypersurface $S_t = \{z\in M: t(z)=t\}$ is a  Cauchy hypersurface.

(b) Let $\gamma_x: \R \rightarrow M$ be the curve in $M$ characterized by:
$$ \Psi (\gamma_x (t)) = (t, x) , \quad \quad \forall t \in \R .$$
Then the continuous curve $\gamma_x$ is timelike in the following sense:
$$ t < t' \Rightarrow \gamma_x(t) << \gamma_x(t').$$ 
\end{prop} 
\noindent 
{\it Sketch of proof.} Define $t(z)$  as in (\ref{tiempo}). By Proposition \ref{l0} 
$S:= t^{-1}(0)$ is a Cauchy hypersurface, and one only has to choose $\rho$ from $\Phi^{-1}$ as in the proof of Lemma \ref{ro}. \cvd

\noindent
Notice that, if function $t$ were smooth, then the property (b) 
would imply directly that $\nabla t$ is everywhere causal. Even more, if $\gamma$ were an integral curve of $\nabla t$ and $\gamma'(t_0)$ were lightlike, $\gamma'$ could not be lightlike on some open interval containing $t_0$ (as $t$ is taken from Proposition \ref{l0}, its item (2) holds).

Notice that Lemma \ref{ro} and Proposition \ref{p0} give two types of topological splittings $\Phi, \Psi$ for $M$, being the curve $s \rightarrow (s,x)$  timelike and smooth for $\Phi$, and the hypersurface $t=$constant a Cauchy hypersurface for $\Psi$.
In what follows, the properties of the Cauchy hypersurfaces of a globally hyperbolic spacetime $M$ will be studied. Then,  we will assume that a {\em topological splitting} either as in Lemma \ref{ro} or as in Proposition \ref{p0} is fixed, and we will drop $\Phi , \Psi$ writting simply $M=\R \times S$.

\section{Some properties of Cauchy hypersurfaces} \label{s3}

First, recall the following technical result. 

\begin{lema} \label{00} Let $M$ be a spacetime and $N$ a closed connected spacelike hypersurface.

(1) A closed curve that meets $N$ exactly once and then transversally is not freely homotopic to a closed curve that does not meet $N$.

(2) If $N$ separates $M$ (i.e., $M\backslash N$ is not connected) then $N$ is achronal.
\end{lema}
Assertion (1) comes from
intersection theory; it is a  particular case of, for example, 
\cite[Sect. 2.4, Theorem in p. 78]{GP}. 
The proof of (2) can be seen in \cite[Lemma 14.45(2)]{O}.

\begin{prop} \label{l1}
Let $M$ be a spacetime which admits a Cauchy hypersurface $S$. Then any closed connected spacelike hypersurface $N$ which does not intersect $S$ is achronal.
\end{prop}

\dem By Lemma \ref{00}(2) it is enough to show that $N$ separates $M$. Otherwise, reasoning by contradiction, recall that there exists a closed curve $\gamma: [-1,1] \rightarrow M$ which intersects $N$ exactly once and, then, transversally. In fact, it is enough to take $\gamma$ on some closed interval $[-\epsilon, \epsilon], \epsilon >0$  transversal to $N$. Then, use that $M\backslash N$ is open and connected (thus, connected by arcs) to join $\gamma(-\epsilon)$ and $\gamma(\epsilon)$.   

Consider the functions $\Phi$ and $\rho: M  \rightarrow S$ in the proof of Lemma \ref{ro}.
 In order to obtain a contradiction 
with Lemma \ref{00}(1), it is enough to prove that $\gamma$ and 
$\rho \circ \gamma$  are freely homotopic. As the homeomorphism $\Phi$ is constructed from
the flow $\phi$ of vector field $T$, there exists a continuous function $\mu$ such that $\rho \circ \gamma(t) = \phi_{\mu(t)}(\gamma(t))$ for all $t\in [-1,1]$.
Then 
$$H(s,t)= \phi_{s \mu(t)}(\gamma(t)) \quad \quad \forall (s,t) \in [0,1]\times [-1,1]$$
is the required free homotopy. 
\cvd 
\smallskip
\noindent 

\begin{rema} \label{re1}
{\rm
Neither the hypothesis ``closed'' nor the hypothesis on the intersection ($N \cap S = \emptyset$) can be dropped, as one can check easily. In fact, consider a cylinder
$\R\times \S^1$ ($\S^1= \R/2\pi\Z$), endowed with the Lorentzian metric 
$-dt^2+ d\theta^2$, which admits the Cauchy hypersurface 
$S=\{-1\} \times \S^1$. For positive slope $c<1$, the curve $t(\theta)= c \, \theta$ corresponds with a helix $H$, which is a non-achronal complete spacelike closed hypersurface and crosses $S$. If the constant $c$ is replaced by a  function $ c(\theta)$ with $0< \dot c(\theta)\theta + c(\theta) <1$ and lim$_{x\rightarrow \pm\infty} c(\theta) = \pm 1$ then the corresponding helix $H'$ does not crosses $S$ (in fact, it lies strictly between two Cauchy hypersurfaces). But $H'$ fails to be closed and is neither achronal.
}
\end{rema}
Recall that, as $N$ is achronal, then it can be seen as a graph on an open subset of the Cauchy hypersurface. In order to assume that this graph is defined on all the hypersurface, the following result gives a sufficient condition.

\begin{prop} \label{l2} Let $M$ be a globally hyperbolic spacetime and $S_1$, $S_2$ be two disjoint Cauchy hypersurfaces, with $S_1 \subset J^-(S_2)$,  that is, $U= I^+(S_1) \cap I^-(S_2) \neq \emptyset$. Consider the topological splitting $M = \R \times S_1$ in Lemma \ref{ro}. 

Then, any closed connected spacelike hypersurface $S\subset U=I^+(S_1) \cap I^-(S_2)$ is a graph on 
all $S_1$, i.e.: there exist a continuous function $\lambda: S_1 \rightarrow (0,\infty)$ such that
$ S= \{ (\lambda(x),x): x \in S_1 \}. $
\end{prop}
\dem First, let us see that the Cauchy hypersurface $S_2$ is a graph too. Consider the
canonical projections
$$\rho : \R \times S_1 \rightarrow S_1 \; , \quad  \quad 
\pi_{R}: \R \times S_1 \rightarrow \R,$$ where we have written $\rho \equiv \pi_{S_1}$ consistently with (\ref{defro}). By the construction of the topological splitting $\Phi$ in Lemma \ref{ro} from the (timelike) integral curves of $T$, it is clear that $\rho|_{S_2}: S_2 \rightarrow S_1$ is continuous and injective. Interchanging the roles of $S_1$ and $S_2$,   $\rho|_{S_2}$ is a homeomorphism and $S_2$ is a graph:
 $$ S_2= \{ (\rho|_{S_2})^{-1}(x): x \in S_1 \} = \{ (\lambda_2(x),x): x \in S_1 \},$$
where $\lambda_2(x) = \pi_{R}\circ (\rho|_{S_2})^{-1}(x)$ for all $x\in S_1$.

For the hypersurface $S$, as it is achronal (Proposition \ref{l1}), the restriction
$\rho|_S: S \rightarrow S_1$ is continuous and injective. Thus, from the theorem of the invariance of the domain, $\rho|_S$ yields a homeomorphism between $S$ and an open subset $U_1:= \rho(S) \; \subseteq S_1$. Therefore, $S$ is a graph on $U_1$:
$$ S= \{ (\rho|_S)^{-1}(x): x \in U_1 \} = \{ (\lambda(x),x): x \in U_1 \},$$
for some continuous function $\lambda$ on $U_1$. We only must prove that $U_1$ is closed in $S_1$. Previously, 
recall that, from the construction of the topological splitting $\R \times S_1$ and the inclusion $S\subset I^+(S_1) \cap I^-(S_2)$:
\be \label{ec}
x << (\lambda(x),x) << (\lambda_2(x),x) , \quad \quad \forall x \in U_1.
\ee

Now, let $\{x_n\}_n$ be a sequence in $U_1 \subset S_1$ which converges to a point $x_0$ in the closure of $U_1$. In order to prove that $x_0\in U_1$, let
$$K_1 = \{ x_n: n \in \N \} \cup \{x_0\} \; \subset S_1,$$
$$K_2 = \{ (\lambda_2(x_n), x_n): n\in \N \} \cup \{(\lambda_2(x_0), x_0)\} \; \subset S_2.$$ 
and define $K= J^+(K_1) \cap J^-(K_2)$, which is compact (see (\ref{jk})). From (\ref{ec}), $K$ contains the sequence
\be \label{es}
\{(\lambda(x_n), x_n): n\in \N\} \; \subset S. 
\ee
As $S$ is closed, this sequence is contained in the compact subset $K_S= K\cap S$. Thus, it converges, up to a subsequence, to a point $(\lambda_0', x_0')$ of $S$. But
$$ x_0' = \pi(\lambda_0', x_0')= \pi(\lim_n (\lambda(x_n), x_n)) = \lim_n x_n = x_0.$$
That is,  $x_0 = x_0' \in \pi(S)=U_1$, as required. \cvd

\begin{rema} \label{re2} {\rm
Even though, by Proposition \ref{l1}, only one of the inclusions
$S\subset I^+(S_1)$ or $S\subset I^-(S_2)$ is enough to ensure the achronality of $S$,  both inclusions are needed for Proposition \ref{l2}. The reason relies in the central role of inequality (\ref{ec}), and it is not difficult to obtain a counterexample if one of them is removed. In fact, consider on $\R^2$ the warped metric $g= -dt^2+ f(t)^2dx^2$ with $f>0$ and $\partial_t$ future-directed. Each hypersurface  $t=$constant is a Cauchy hypersurface, because the spacetime is conformal to $(I\times \R, g^*=-ds^2 + dx^2)$, where $I$ is some interval of $\R$ and $ds=dt/f$. Now, the graph $S$ 
of a smooth curve $x\rightarrow t(x)$ is a spacelike hypersurface if and only if
$|dt/dx| < f(t(x))$. But $f$ can grow fast enough in such a way that the inextendible domain of $t(x)$ is a finite interval, and $S$ will be closed but not a graph on all $S_1$.

As a concrete example, consider $f(t) = \cosh t$ for all $t$ (the spacetime is then isometric to the universal covering of 2-dimensional de Sitter spacetime), choose $S_1$ 
as the hypersurface $t\equiv -1$
and put $t(x) = {\rm tg}^2(x/4)$, which satisfies the required inequality
$|dt/dx| = \sqrt t(1+t)(x) /2< f(t(x))$. 
Recall that the hypersurface $S=\{({\rm tg}^2(x/4), x): \, x\in (-2\pi, 2\pi )\}$, which lies in $I^+(S_1)$, is not only closed but also complete (the $g$-length of the graph $(t(x),x)$ restricted to both $x\in (-2\pi,0)$ and $x\in (0,2\pi)$ is infinity). Finally, notice that this example can be easily modified to make $S_1$ compact (take the  quotient cylinder
generated from the isometry $(t,x) \rightarrow (t,x+4\pi)$).
}
\end{rema}
Thus, as  a straightforward consequence we obtain the following result (an alternative proof can be found in 
\cite[Corollary 2]{Ga}):
\begin{coro} \label{p1} Let  $S_1$ and $S_2$ be two disjoint Cauchy hypersurfaces of $M$ with $S_1 \subset I^-(S_2)$. Then:

Any closed connected spacelike hypersurface $S$ contained in $U= I^+(S_1) \cap I^-(S_2)$ is a Cauchy hypersurface.
\end{coro}
\dem From Lemma \ref{l1},  $S$ is achronal, and we only must check that each inextendible timelike curve $\gamma$ crosses it. 
As $S$ is a graph on $S_1$, it separates  $\R \times S_1$ in two disjoint open subsets: 
$W^-=\{(t,x): t<\lambda(x)\}$, which includes $S_1$, and  
$W^+=\{(t,x): t>\lambda(x)\}$, which includes $S_2$. As $\gamma$ crosses the Cauchy surfaces $S_1$ and $S_2$, it will cross $S$ too. \cvd

\section{Smooth Cauchy hypersurfaces} \label{s4}

In what follows we will use convex open subsets of $M$. Recall that an open set is called {\em convex} if it is a (starshaped) normal neighborhood of all its points; every point $p\in M$  has a convex neighborhood ${\cal C}_p$ (${\cal C}_p$ can be also chosen {\it simple} in the terminology of \cite{Pe}, \cite{D}, i.e. convex with compact closure included in a bigger convex set); some properties of this sets in relation to causality can be seen in \cite[14.2]{O}. When a convex set ${\cal C}$ is regarded as a spacetime, the {\em time--separation} or Lorentzian distance on ${\cal C}$ has especially good properties. In particular, it is not only continuous but also smooth whenever it does not vanish (see, for example, \cite[Lemmas 5.9 and 14.2(1)]{O}).

\begin{lema} \label{local}
Let $M$ be a globally hyperbolic spacetime with a topological splitting $\R \times S$ as in Proposition \ref{p0}. Let $t_1<t_2$ and denote the corresponding Cauchy hypersurfaces $S_{t_1} \equiv S_1$, $S_{t_2} \equiv S_2$. Fix
$p \in S_{2}$, and a convex neighborhood of $p$, ${\cal C}_p \subset
I^+(S_{1})$. 

Then there exists a smooth function
$$
h_p: M \rightarrow [0, \infty)
$$
which satisfies:

(i) $h_p(p)=1$.

(ii) The support of $h_p$ (i.e., the closure of $h_p^{-1}(0, \infty)$)  is compact and included in ${\cal C}_p \cap I^+(S_1)$.

(iii) If $q \in J^-(S_{2}) $ and $h_p(q) \neq 0$ then $\nabla h_p(q)$ is timelike and past-pointing.
\end{lema}

\dem Choose $p' \in I^-(p)\cap I^+(S_1)$ such that $J^+(p') \cap J^-(S_{2}) \subset {\cal C}_p$ and define $h_p$ on 
$I^-(S_{2})$ as the $C^k$ function:
\be \label{hp}
h_p(q) = 
\left\{
\begin{array}{ll}
\tau (p',p)^{-2k} \; \cdot \, \tau (p',q)^{2k} & \hbox{if} \; k<\infty \\
e^{\tau (p',p)^{-2}} \; \cdot \, e^{-\tau (p',q)^{-2}} & \hbox{if} \; k=\infty 
\end{array}
\right.
\ee
where $\tau$ is the time-separation on ${\cal C}_p$ regarded as a spacetime
($h_p$ is regarded as 0 on $I^-(S_{2})\backslash  {\cal C}_p$). Now, construct any $C^k$ extension of $h_p$ out of $I^-(S_{2})$ 
such that the support of $h_p$ is included in ${\cal C}_p$ and $h_p\geq 0$.\footnote{This extension can be easily carried out because the support of $h_p$ on $J^-(S_{2})$ 
lies in the compact subset $K=J^+(p') \cap J^-(S_{2})\subset {\cal C}_p$. Fix an open neighbourhood $U$ of $K$ included in $ {\cal C}_p$, consider the covering $(U, {\cal C}_p\backslash K)$ of ${\cal C}_p$, and take the subordinate partition of the unity $\{ \mu, \mu'\}$ with support$(\mu ) \subset U$ . Then, $h_p$ can be chosen on all ${\cal C}_p$ as the expression in (\ref{hp}) multiplied by  $\mu$.} Obviously, conditions (i) and (ii) are fulfilled and, in order to check (iii), take into account that the gradient of $-\tau (p', \cdot)^2/2$ at $q$ is the velocity of the unique (timelike, future-pointing) geodesic $\sigma:[0,1] \rightarrow {\cal C}_p$ from  $p'$ to $q$. 
\cvd

\noindent Recall that the existence of a Lorentzian metric on $M$ implies its paracompactness, i.e., for any open covering ${\cal W}$ of $M$ there exists a {\em locally finite {\bf refinement}} 
\cite[Vol II, Addendum 1]{Sp}. We will need a technical property related to paracompactness. Essentially, take any complete distance $d_R$ on $M$ compatible with its topology, and assume that there exist a $d_R$-bound of the diameter of the elements of the covering ${\cal W}$. Then there exist a {\em locally finite {\bf subcovering} ${\cal W'}$} of ${\cal W}$. 
More technically, we will need:
\begin{lema} \label{topol}
Let $d_R$ be the distance on $M$ associated to any auxiliary complete Riemannian metric $g_R$.  Let $S$ be a closed subset of $M$ and ${\cal W}= \{ W_\alpha, \alpha\in {\cal I} \}$ a collection of open subsets of $M$ which cover $S$. Assume that each $W_\alpha$ is included in an open subset ${\cal C}_\alpha$ and the $d_R$-diameter of each ${\cal C}_\alpha$ is smaller than 1. Then there exist a subcollection ${\cal W}'=\{ W_j: j\in \N\} \subset {\cal W}$ which covers $S$ and is locally finite (i.e., for each $p\in \cup_j W_j$ there exists a neighborhood $V$ such that $V\cap W_j = \emptyset$ for all $j$ 
but a  finite set of indexes). Moreover, the collection $\{ {\cal C}_j: j\in \N\}$ (where each $W_j \in {\cal W}'$ is included in the corresponding ${\cal C}_j$) is locally finite too.
\end{lema}
\dem Fix $p\in M$ and consider the open and closed balls, resp., $B_p(r), \bar B_p(r)$ of center $p$ and radius $r>0$ for the distance $d_R$. As each closed ball is compact, the following subsets  are compact too:
$$ K_m= \bar B_p(m)\backslash B_p(m-1), \quad \quad  \quad \quad (M\subset \cup_mK_m ).$$
$$ S_m= K_m\cap S, \quad \quad \quad \quad (S\subset \cup_mS_m ),$$
for all  $m\in \N$. 
From the compactness of $S_m$, a finite subset $\{W_{1m}, \dots W_{k_mm}\} \subset {\cal W}$ covers $S_m$. Then, take:
$$ {\cal W}'= \{W_{km} : m\in \N ,\; k= 1,\dots k_m \}.$$
Clearly, ${\cal W}'$ covers $S$, and ${\cal W}'$ (as well as the corresponding collection of the ${\cal C}_j$'s) is locally finite because, if $|m-m'|\geq 3$, then
$ W_{km} \cap W_{k'm'} = \emptyset$ (use the definition of the $K_m's, S_m$'s and the inequality
diam$\, W_{\alpha}<1$, for all $\alpha$).
\cvd

\noindent 
\begin{prop} \label{global}
Let $M$ be a globally hyperbolic spacetime with topological splitting $\R \times S$ and fix
$t_1<t_2$, $S_i$, as in Lemma \ref{local}.

Then there exists a smooth function
$$
h: M \rightarrow [0, \infty)
$$
which satisfies:

\begin{itemize}
\item[1. ] $h(t,x)=0$ if $t\leq t_1$.
\item[2. ]  $h(t,x) > 1/2$ if $t = t_2$. 
\item[3. ] The gradient of $h$ is timelike and past-pointing on the open subset 
$V=h^{-1}((0,1/2)) \cap I^-(S_{2})$.

\end{itemize}
As a consequence, $S^h_s= h^{-1}(s) \cap J^-(S_{2})$ is a closed smooth spacelike hypersurface which lies in $I^+(S_{1})\cap I^-(S_{2})$, for each $s\in (0,1/2)$.
\end{prop}
\dem Fixed a complete auxiliary distance $d_R$ as in Lemma \ref{topol}, take, for any $p\in S_{2}$, a convex neighborhood ${\cal C}_p$ with diameter smaller than 1 and the corresponding function $h_p$ as in Lemma \ref{local}. Let 
$W_p = h_p^{-1}\left((1/2,\infty)\right)$ ($W_p \subset {\cal C}_p$). Obviously, 
$${\cal W} = \{ W_p, \; p\in  S_{2} \} $$
covers the closed hypersurface $S_{2}$ and lies in the hypothesis of Lemma \ref{topol}. 
Now, if ${\cal W}'= \{W_i ; \; i\in \N \}$ is the locally finite subcovering given by this 
lemma, and for each $W_i (\equiv W_{p_i})$,  $h_i$ is the corresponding function with support ${\cal C}_i$ ($W_i \subset {\cal C}_i$), we put:
$$ h = \sum_i h_i , $$
which is well-defined and smooth because of the local finiteness of the supports of the $h_i$ 's. Clearly, $h$ satisfies the  items 1, 3 because each $h_i$ satisfies them, and item 2 is satisfied because each $h_i(p)> 1/2$ for any $p\in W_i$, and $ S_{2} \subset \cup_i W_i$.

To prove the last assertion, notice that, by item 3, any $s\in (0,1/2)$ is a regular value of the restriction of $h$ to $I^-(S_{2})$; thus, $S^h_s$ is a spacelike hypersurface, and it is also closed in $I^-(S_{2})$. 
Nevertheless, the closure of $I^-(S_{2})$ is $S_{2}$  and, by item 2, no frontier point of $I^-(S_{2})$ can be a frontier point of $S^h_s$. Thus, $S_s^h$ is closed in $M$, as required.
\cvd

\noindent {\it Proof of Theorem \ref{tsuave}}. To obtain the smooth hypersurface, apply Corollary \ref{p1} to any hypersurface $S^h_s$ yielded by Proposition \ref{global}. For the diffeomorphism, use Lemma \ref{ro}. \cvd

\end{document}